\documentclass[tightenlines,twocolumn,prl,notitlepage,nofootinbib]{revtex4-1}

\usepackage{graphicx}
\usepackage{amsmath}
\usepackage{amsfonts}
\usepackage{amssymb}
\usepackage{float}
\usepackage{hyperref}
\usepackage{multirow}
\usepackage{bbold}
\usepackage{xcolor}
\usepackage{soul}
\usepackage{makecell}
\usepackage{minibox}

\begin{document}

\pagestyle{plain}


\title{Neutron capture and the antineutrino yield from nuclear reactors}

\author{Patrick Huber}
\email{pahuber@vt.edu}

\author{Patrick Jaffke}
\email{pjaffke@vt.edu}

\affiliation{Center for Neutrino Physics,
  Virginia Tech, Blacksburg, VA 24061, USA}

\date{\today}

\begin{abstract}
We identify a new, flux-dependent correction to the antineutrino
spectrum as produced in nuclear reactors. The abundance of certain
nuclides, whose decay chains produce antineutrinos above the threshold
for inverse beta decay, has a nonlinear dependence on the neutron
flux, unlike the vast majority of antineutrino producing nuclides,
whose decay rate is directly related to the fission rate. We have
identified four of these so-called nonlinear nuclides and determined
that they result in an antineutrino excess at low-energies below
$3.2\,\mathrm{MeV}$, dependent on the reactor thermal neutron flux.
We develop an analytic model for the size of the correction and
compare it to the results of detailed reactor simulations for various
real existing reactors, spanning 3 orders of magnitude in neutron
flux. In a typical pressurized water reactor the resulting correction
can reach $\sim0.9$\% of the low energy flux which is comparable in size to other, known
low-energy corrections from spent nuclear fuel and the non-equilibrium
correction. For naval reactors the non-linear correction may reach the 10\% level.
\end{abstract}

\maketitle


Science with antineutrinos and nuclear reactors and have been
intimately connected since the discovery of the antineutrino by Cowan
and Reines~\cite{Cowan:1992xc}. Reactors are the largest terrestrial
source of antineutrinos through the production of unstable fission
fragments. These fission fragments are neutron-rich and, thus, will
beta decay to stability producing antineutrinos. An average of six
beta decays occurs per fission, thus a $1\,\mathrm{GW_{th}}$ reactor
will produce $\mathcal{O}(10^{20})\,\bar\nu /\mathrm{sec}$. The vast
majority of reactor nuclides lighter than uranium are generated directly
as a fission product or by the beta decays of fission products, for instance
\begin{equation}
^{99}\mathrm{Zr} \rightarrow ^{99}\mathrm{Nb} \rightarrow ^{99}\mathrm{Mo} \rightarrow ^{99}\mathrm{Tc}
\label{eq:betachain}
\end{equation}
where we have truncated the chain at $^{99}$Tc as its half-life of
$\sim2\times10^5\,\mathrm{y}$ allows us to consider it stable. Here,
the daughter nuclides are produced from decays of their parents, which
are dominantly produced via fissions. Thus, the decay rates of both the
daughters and parents in the chain are linearly dependent on the 
fission rates. Equivalently, these nuclides are said to be linear in the
neutron flux $\phi$ as the fission rate goes as $\phi\Sigma_\mathrm{fiss}$ for
a macroscopic fission cross-section $\Sigma_\mathrm{fiss}$. A second mechanism
for antineutrino production is from neutron captures on certain isotopes,
such as:
\begin{equation}
^{99}\mathrm{Tc} + \mathrm{n} \rightarrow ^{100}\mathrm{Tc}
\label{eq:ncapchain}
\end{equation}
where the neutrons are the prompt neutrons from fission,
thermalized by the moderator. Nuclides that are primarily produced via
neutron captures require two neutrons: one to initiate the fission
whose fission products result in  a beta decay chain yielding the capture
isotope, $^{99}$Tc in above example, and a second neutron for the
actual neutron capture. Thus naively, one would conclude that the
production of $^{100}$Tc is \emph{quadratic} in the neutron
flux. Thus, these nuclides will be produced in different quantities
for different values of $\phi$ even if $\phi\,T_{\mathrm{irr}}$ is kept
constant; we therefore name these nonlinear nuclides. This letter
examines how many such nonlinear nuclides with a relevant
antineutrino yield exist and how large the resulting correction to the
antineutrino spectrum can become.

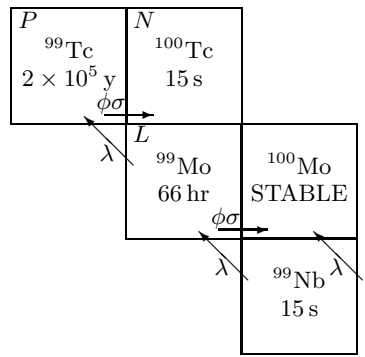
\begin{figure}[h!]
\setlength{\unitlength}{0.08in}
\centering
\hspace{-10mm}
\vspace{-5mm}
\begin{picture}(25,25)
\put(2.75,18.25){\framebox(7.5,7.5){\minibox[c]{$^{99}$Tc \\ $2\times10^5\,\mathrm{y}$}}}
\put(3.2,24.5){$P$}
\put(10.25,10.7){\framebox(7.5,7.5){\minibox[c]{$^{99}$Mo \\ $66\,\mathrm{hr}$}}}
\put(10.7,17.){$L$}
\put(17.8,3.2){\framebox(7.5,7.5){\minibox[c]{$^{99}$Nb \\ $15\,\mathrm{s}$}}}
\put(10.25,18.25){\framebox(7.5,7.5){\minibox[c]{$^{100}$Tc \\ $15\,\mathrm{s}$}}}
\put(10.7,24.5){$N$}
\put(17.8,10.7){\framebox(7.5,7.5){\minibox[c]{$^{100}$Mo \\ $\mathrm{STABLE}$}}}
\put(10.7,15.5){\vector(-1,1){3.2}}\put(8.5,15.8){$\lambda$}
\put(18.2,8.){\vector(-1,1){3.2}}\put(16,8.3){$\lambda$}
\put(25.7,8.){\vector(-1,1){3.2}}\put(23.5,8.3){$\lambda$}
\put(8.8,18.8){\vector(1,0){3.2}}\put(8.3,19.2){$\phi\sigma$}
\put(16.3,11.3){\vector(1,0){3.2}}\put(15.8,11.7){$\phi\sigma$}
\end{picture}
\caption{\label{fig:NonLinearNuclide} Example of a typical nonlinear
  beta-decaying nuclide ($N$), $^{100}$Tc.
Half-lives taken from ENSDF~\cite{ENSDF}.}
\end{figure}

To be a relevant nonlinear nuclide $N$, several conditions have to be met:
\begin{enumerate}
\item Large cumulative fission yield, $Z_P$ of the capture isotope $P$.
\item Large neutron capture cross section $\sigma^c_P$.
\item The nonlinear nuclide must decay sufficiently quickly, that is the decay constant $\lambda$ must be large enough.
\item The beta decay of the nonlinear nuclide has to have an endpoint
  above the inverse beta decay threshold of 1.8\,MeV.
\end{enumerate}
There are approximately $20$ candidate nuclides fulfilling these
conditions.  An example of a relevant nonlinear nuclide, $^{100}$Tc,
is given in Fig.~\ref{fig:NonLinearNuclide}.

Here, $^{100}$Tc is the beta decaying nonlinear nuclide $N$ and it is
fission-blocked from the beta decay chain by its stable isobar
$^{100}$Mo. Thus, $^{100}$Tc is practically absent from fission
products, {\it i.e.} its fission yield $Y_f$ is negligible. Being
fission blocked by a double-beta decay isotope, like $^{100}$Mo, is
characteristic for all candidates. The $N$ production is then
primarily governed by its precursor nuclide $P$, in this case
$^{99}$Tc. Furthermore, $^{99}$Tc is relatively stable and linear as
it is fed through its own decay chain meaning that, with a large
enough cumulative yield and neutron capture cross-section, the
production of $^{100}$Tc may be non-negligible.  To simplify our
discussion we consider only $N$ that have stable precursors $P$
(including $^{100}$Tc), are significantly blocked ($Y_N^f\ll1$), and
have a significant feeder cumulative yield $(\sum_{f=1}^{N_f}Z_P^f\geq
0.025)$, where we used the JEFF-3.1~\cite{JEFF} fission yields. With
these criteria we are able to reduce our list to four major nonlinear
nuclides listed in Tab.~\ref{tab:NonLinears}. Note, that $^{100}$Tc,
$^{104}$Rh, $^{110}$Ag show predominantly ($>95\%$) allowed Gamow-Teller
decays, whereas $^{142}$Pr exhibits a non-unique forbidden decay, which as
we will see later, contributes less than 10\% to the total
nonlinear correction.
\begin{table}[t]
\centering
\renewcommand{\arraystretch}{1.25}
    \begin{tabular}{|c|c||c|c|c|c|}
        \hline 
	\multicolumn{2}{|l||}{} & $^{100}$Tc & $^{104}$Rh & $^{110}$Ag & $^{142}$Pr \\ \hline \hline
	\multicolumn{2}{|l||}{$N$ \hspace{2mm} $E_0\,\,\,\mathrm{(MeV)}$} & 3.2 & 2.4 & 2.9 & 2.2 \\ \hline
	\multicolumn{2}{|l||}{$N$ \hspace{2mm} $\tau_{1/2}\,\,\,\mathrm{(sec)}$} & 15.5 & 42.3 & 24.6 & 68830 \\ \hline \hline
	\multirow{3}{*}{\makecell{$P$ Cumul. \\ Fission Yields \\ (atoms/fiss.)}} & $^{235}$U & 0.061 & 0.031 & 0.00029 & 0.059 \\ \cline{2-6}
	& $^{239}$Pu & 0.062 & 0.069 & 0.017 & 0.052 \\ \cline{2-6}
	& $^{241}$Pu & 0.056 & 0.065 & 0.030 & 0.049 \\ \hline
	\multicolumn{2}{|l||}{$P$ \hspace{2mm} $\sigma^c_P\,\,\,\mathrm{(b)}$} & 17.0 & 127 & 80.9 & 6.53 \\ \hline \hline
	\multicolumn{2}{|l||}{$L$ \hspace{2mm} $\tau_{1/2}^L\,\,\,\mathrm{(d)}$} & 2.75 & 39.3 & 0.57 & 32.5 \\ \hline
	\multicolumn{2}{|l||}{$L$ \hspace{2mm} $\sigma^c_L\,\,\,\mathrm{(b)}$} & 1.57 & 7.08 & 18.2 & 26.7 \\ \hline
    \end{tabular}
\caption{\label{tab:NonLinears} Properties of the four selected
  nonlinear nuclides (N) including their beta endpoints (MeV),
  half-lives (sec), cumulative precursor (P) fission yields
  (atoms/fission), and their precursor flux-averaged thermal neutron
  capture cross-section (b) taking the thermal flux from Fig. 3 of
  Ref.~\cite{DOEHNDBK} and the cross-sections from
  CINDER~\cite{CINDER}. Also provided are the long-lived feeder parent
  (L) neutron capture cross-sections and half-lives.}
\end{table}

From Fig.~\ref{fig:NonLinearNuclide} it is apparent we must solve a
set of three linearly coupled non-homogeneous differential equations,
the Bateman equations~\cite{Bateman:1910}, in order to express the
abundance of N in terms of the thermal neutron flux $\phi$
and the irradiation time $T_\mathrm{irr}$. Similar sets have been solved, without
the neutron component, as an eigenvalue problem~\cite{Moral:2003} and
recursively~\cite{Cetnar2006640}.

The limiting cases of the solutions can be identified from the
information provided in Tab.~\ref{tab:NonLinears}. For irradiation
times larger than the longest involved half-lives, which generally
occur for the long-lived precursor parent $L$, and the relevant
half-lives range from 0.57\,d to 39.3\,d, we can assume that the
isotope $L$ is in equilibrium. 

Once $L$ has reached equilibrium, the next concern is that capture
directly from the long-lived nuclide to a stable isotope can prevent
the production of the neutron capture isotope. The decay rate $\ln
2/\tau_{1/2}$ of the long-lived nuclide equals the capture rate for a
neutron flux density $\tilde \phi$ of
\begin{equation}
\tilde \phi=\frac{\ln 2}{{\tau_L}_{1/2}}{\sigma^c_L}\,,
\end{equation}
using the values given in Tab.~\ref{tab:NonLinears} this yields a
range of
$\tilde\phi=9\times10^{15}-2\times10^{18}\,\mathrm{s}^{-1}\,\mathrm{cm}^{-2}$,
which is nearly an order of magnitude above the values found for any
of the reactors considered here. Thus we conclude, that this mechanism
can practically be neglected. Note, that for $^{136}$Cs, the
neutron capture on $^{135}$Xe with a cross section of $2.7\,\mathrm{Mb}$
prevents any significant production.

The decay rate of the antineutrino producing nonlinear nuclide is large
with half-lives in seconds to hours range and thus will be always in
equilibrium with its much slower production rate. Therefore, for
irradation times which are long compared to the half-lives of $L$,
the production rate of the nonlinear nuclide, which is the
same as its decay rate is given by
\begin{equation}
\label{eq:rate}
\Gamma_{nonlinear}\propto\underbrace{\Sigma_{\mathrm{fiss}} \phi Z_P  T_\mathrm{irr}}_{\text{atoms of P}} \sigma_P^c \phi\propto T_\mathrm{irr} \phi^2\,,
\end{equation}
hence the name nonlinear nuclide. The decay rate of a fission product
in equilibrium is given proportional to $\phi$ and thus the relative
contribution of a nonlinear nuclide scales as $T_\mathrm{irr}
\phi$. From Eq.~\ref{eq:rate} and Tab.~\ref{tab:NonLinears} we also
can conclude that $^{104}$Rh will have the largest contribution for
the fissile isotopes investigated, for fission of $^{235}$U the second
most important nonlinear nuclide is $^{100}$Tc, whereas for the
fission of both plutonium isotopes the second largest contribution
stems from $^{110}$Ag. We note, that reactors with a high neutron flux
density and very long core life-times, in principle, can
exhibit corrections in the 5-10\% range. In one example assuming a
neutron flux density of $10^{15}\,\mathrm{cm}^{-2}\,\mathrm{s}^{-1}$
and after an irradiation of 5 years, we find, based on our analytic
calculation, a 4\% correction. Clearly, naval reactors fulfill these
characteristics and a precise measurement of the nonlinear correction
to their antineutrino emissions may allow to draw conclusions about
some of the design characteristics, like core size, and operational
history.

\begin{table*}[tb]
\centering
\renewcommand{\arraystretch}{1.25}
    \begin{tabular}{|c||c|c|c|c|c|c|c|c|c|}
        \hline
	\multirow{2}{*}{} & \multirow{2}{*}{$5\,\mathrm{MW_e}$} & \multirow{2}{*}{IR40} & \multicolumn{2}{c|}{PWR} & \multirow{2}{*}{IRT} & \multicolumn{3}{c|}{ILL} & \multirow{2}{*}{HFIR} \\ \cline{4-5} \cline{7-9}
	& & & 1-batch & 3-batch & & $^{235}$U & $^{239}$Pu & $^{241}$Pu & \\ \hline
	Fuel/Moderator & NU+C & NU+D$_2$O & \multicolumn{2}{c|}{LEU+H$_2$O} & HEU+H$_2$O & \multicolumn{3}{c|}{HEU+D$_2$O} & HEU+H$_2$O \\ \hline \hline
	Burn-up [MWd/t] & $32380$ & $31200$ & $31510$ & $1890000$ & $2230$ & $7.3\times10^{-5}$ & $1.1\times10^{-4}$ & $1.7\times10^{-4}$ & $2550$ \\ \hline
         $\phi\,\mathrm{[n/cm^2/sec]}$ & $1.6\times10^{12}$ & $3.6\times10^{13}$ & $4.4\times10^{13}$ & $4.4\times10^{13}$ & $1.5\times10^{14}$ & $3.3\times10^{14}$ & $3.3\times10^{14}$ & $3.3\times10^{14}$ & $2.5\times10^{15}$ \\ \hline
	Max[$\langle\Phi_{\mathrm{NL}}/\Phi_{\mathrm{R}}\rangle_T$] [\%] & $0.027$ & $0.15$ & $0.24$ & $0.92$ & $0.11$ & $3.1\times10^{-5}$ & $2.6\times10^{-3}$ & $4.7\times10^{-3}$ & $0.10$ \\ \hline
    \end{tabular}
\caption{\label{tab:Rxspecs} Details of the reactor calculations via
  Origen including the burn-up, thermal neutron flux, and design
  details. The burn-up calculations for the PWR
  cores and IRT have been separated into their individual batches and
  elements, respectively.}
\end{table*}

We can formulate an explicit solution for the nonlinear nuclides by
solving the corresponding set of Bateman equations. The long-lived
nuclide is dominantly produced via fission and destroyed through its
neutron captures and decays.  The precursor is produced via fission
and decays from $L$. It is destroyed by neutron captures. Finally, the
nonlinear nuclide is produced solely through captures on $P$ and is
destroyed via its own decays. Therefore, our nonlinear set is given
by:
\begin{equation}
\begin{split}
\frac{dN_L}{dT_\mathrm{irr}} &= \vec{Z}_L\cdot\vec{\mathcal{F}} - \tilde{\lambda}_L N_L \\
\frac{dN_P}{dT_\mathrm{irr}} &= \vec{Y}_P\cdot\vec{\mathcal{F}} + \lambda_LN_L - \phi\sigma^c_PN_P \\
\frac{dN_N}{dT_\mathrm{irr}} &= \phi\sigma^c_PN_P -\lambda_N N_N
\end{split}
\label{eq:NLde}
\end{equation}
where $\tilde{\lambda}_i=\lambda_i+\phi\sigma^c_i$, $\vec{\mathcal{F}}$ is
the fission rate vector, and $\vec{Z}_i$ ($\vec{Y}_i$) is the cumulative
(individual) fission yields. All nuclear parameters are denoted by their subscript.
It is straightforward to solve Eq.~\ref{eq:NLde} analytically, but the
salient features are contained in above description of the limiting
cases. Many of the reactor physics effects neglected in the simplified
reaction network used result in non-negligible corrections and we find
that the analytic result generally is within a factor of two the
solution derived from using a full reaction network. The full reaction
network is evaluated using the Standardized Computer Analyses for
Licensing and Evaluation (SCALE-6.1)~\cite{SCALE} reactor simulation
suite, developed by Oak Ridge National Laboratory.

Now that we have an expression for the activity of these four
nonlinear beta decaying nuclides we can apply a neutrino spectrum to
each decay to generate a neutrino rate. The neutrino spectra are
applied to our four nuclides following Ref.~\cite{Huber:2011wv}, which
generates the neutrino spectra for each nonlinear nuclide. The neutrino
spectra were then summed to determine the total nonlinear correction.

Solving Eq.~\ref{eq:NLde} will lead to an expression for the activity
of the nonlinear nuclides, which can be combined with the spectra of
each nonlinear isotope to produce the total nonlinear spectral
contribution.  Comparing this with the total reactor spectra, shows
that the nonlinear spectra falls off steeply at
$\sim2.4\,\mathrm{MeV}$. This nonlinear spectral contribution is
important as it interferes with other low-energy corrections, such as
the spent fuel signal~\cite{Zhou:2012zzc}, the non-equilibrium
correction for inverse beta decay experiments\footnote{For elastic
  antineutrino-electron scattering experiments at very low energies a
  detailed discussion of the non-equilibrium correction including some
  neutron captures (different than those considered here) can be found
  in Refs.~\cite{Mikaelyan:2002nv,Kopeikin:2001rj} .} which has been
  evaluated in Ref.~\cite{Mueller:2011nm}, where neutron capture was
  specifically neglected.  All of these corrections, including the
  nonlinear correction, will directly impact geoneutrino
  searches~\cite{Sramek:2012hma,Domogatski:2004gs} wherever a sizable reactor
  signal needs to be subtracted, like for instance in JUNO~\cite{Han:2015roa}.
.

Using SCALE, we are able to model nine different reactor configurations,
spanning three orders of magnitude in their thermal neutron flux. The
first is a natural uranium loaded and graphite-moderated reactor,
similar in design to the British Calder-Hall reactor. This reactor is
referred to as the $5\,\mathrm{MW_e}$ reactor and has been previously
modeled~\cite{Christensen:2013eza}. The next reactor uses natural
uranium as fuel and heavy water as a moderator, similar in design to
the CANDU reactors. This reactor, referred to as the IR40, has also
been previously modeled~\cite{PhysRevLett.113.042503}. The third
reactor is fueled with low-enriched uranium (LEU) with a water
moderator. These reactors are pressurized water reactors (PWR)
similar in design to the Daya Bay cores. The Daya Bay reactors have
also been previously modeled to estimate the spent fuel
contribution~\cite{An:2012eh}. The PWR cores are simulated
using a 3-batch method, where a full core consists of 3 parts: a third each
of fresh, once-irradiated, and twice-irradiated fuel. We also include a 
single-batch calculation for comparison. Next, we simulate a research reactor,
named the IRT reactor, which is a pool-type reactor using highly-enriched
uranium (HEU) fuel elements, natural uranium target elements, and water
as a moderator. It was previously simulated, also in 
Ref.~\cite{Christensen:2013eza}. We have also recreated the measurements
conducted at the ILL reactor, irradiating a fissile mass with a specific neutron
flux according to Ref.~\cite{Schreckenbach:1985ep,Haag:2013raa,Hahn:1989zr}.
Finally, we simulate the High Flux Isotope Reactor (HFIR) at Oak Ridge National
Laboratory, which represents the highest steady-state neutron flux
commercially available. Our simulation closely follows that of 
Ref.~\cite{Ilas:2011}. This reactor database spans over $3$ orders of magnitude for the
neutron fluxes and we aim to find a nonlinear correction trend as a
function of $\phi$. Each reactor is irradiated with its own typical
power history using the SCALE simulation suite. The burn-up and reactor
specifications are given in Tab.~\ref{tab:Rxspecs}.

We use the Origen depletion subroutine to compute the fission rates
and nuclide activities as a function of irradiation time. We use the
linear antineutrino yields for $^{235}$U, $^{239}$Pu and $^{241}$Pu
from Ref.~\cite{Huber:2011wv} and for $^{238}$U from
Ref.~\cite{Mueller:2011nm} to convert the fission rates to a total
neutrino spectrum for each reactor during its power cycle. The
nonlinear correction is isolated by selecting the activities of our
four nonlinear nuclides and converting these into a neutrino spectrum
using the beta decay description in Ref.~\cite{Huber:2011wv}. Each
spectrum is binned into $250\,\mathrm{keV}$ bins and a nonlinear
correction is determined from the ratio of the nonlinear contribution
to the total reactor spectrum at all irradiation times. This result is
then used to calculate a time-averaged nonlinear correction.

With this nonlinear low-energy neutrino correction, we can see that
commercial reactors can be very sensitive to the resulting
effects, where it becomes comparable with spent fuel
($\sim 1-2\%$)~\cite{Zhou:2012zzc} and the non-equilibrium
correction ($\sim 1-4\%$)~\cite{Mikaelyan:2002nv,Kopeikin:2001rj}.
Therefore, neutrino experiments will need to consider the nonlinear
correction when predicting the total reactor neutrino spectrum,
especially in the low energy region where detailed reactor
simulations are necessary.

A final item of note is that the widely used measurements of the
cumulative beta spectra from fissions of
$^{235}$U~\cite{Schreckenbach:1985ep}, $^{239}$Pu and
$^{241}$Pu~\cite{Hahn:1989zr}, and now $^{238}$U~\cite{Haag:2013raa}
have utilized research reactors with fluxes of
$\mathcal{O}(10^{14}\,\mathrm{n/cm^2/sec})$. Our analysis has been
conducted to reproduce the measurements conducted by Schreckenbach
{\it et al.} to determine if nonlinear effects appear in these
measurements. A flux of $\phi=3.3\times10^{14}\,\mathrm{n/cm^2/sec}$
was used with irradiation times of $12\,\mathrm{hr}, 36\,\mathrm{hr},
43\,\mathrm{hr}$, and $42\,\mathrm{hr}$ for $^{235}$U, $^{239}$Pu,
$^{241}$Pu, and $^{238}$U, respectively in accordance with
Ref.~\cite{Schreckenbach1981251,VonFeilitzsch:1982jw,Schreckenbach:1985ep,Hahn:1989zr,Haag:2013raa}.
The results for these calculations, shown in Tab.~\ref{tab:Rxspecs}
illustrate that these measurements are not contaminated by nonlinear
corrections and thus the extracted neutrino
fluxes~\cite{Huber:2011wv,Mueller:2011nm} are unaffected.

The lack of nonlinear corrections in the Schreckenbach measurements is
due to the short irradiation times $T_{\mathrm{irr}}$, which are all
less than two days. As we have noted earlier, such as in
Fig.~\ref{fig:NonLinearNuclide}, most of our nonlinear nuclides are
fed via a precursor nuclide $P$ with a long-lived parent $L$. The
large half-lives, relative to $T_{\mathrm{irr}}$, of the $L$ nuclides
prevents the buildup of the feeder nuclides, which then prevents the
buildup of the nonlinear nuclides, thus preserving the Schreckenbach
measurements. This same effect is seen in the diminished nonlinear
correction for the HFIR reactor, which involves irradiation cycles
less than $30\,\mathrm{d}$. Two nonlinear nuclides ($^{104}$Rh and
$^{142}$Pr) are fed through an $L$ with
$\tau_{1/2}\geq30\,\mathrm{d}$, so their contribution to the HFIR
correction is lower than would be expected for longer irradiation
times.

\begin{figure}[tb]
\centering
\includegraphics[width=\columnwidth]{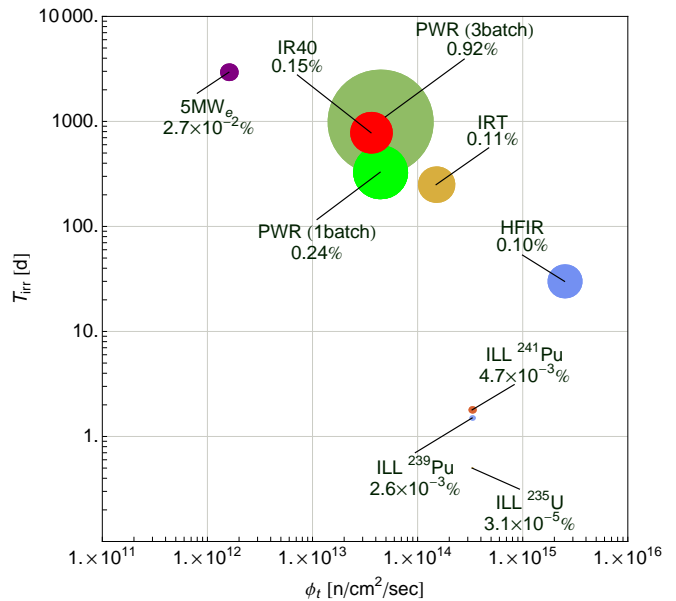}
\caption{\label{fig:NLCorrection} Time-averaged maximum nonlinear
  correction for nine different reactor configurations as computed via
  SCALE. Both a single batch and a 3-batch core were considered for
  the PWR. Area of the disk is proportional to the size of the
  correction.}
\end{figure}


In our note we have introduced a new low-energy correction to the
reactor antineutrino spectrum. This correction is due to nonlinear
nuclides in the reactor, which are dominantly produced via neutron
captures. Demanding that our nuclides of interest meet several
criteria, we have limited the list of these nonlinear nuclides to four
that can impact neutrino studies: $^{100}$Tc, $^{104}$Rh, $^{110}$Ag,
and $^{142}$Pr. We derived an analytic solution for the abundance of
these nuclides in a reactor environment, which depends on the neutron
flux in a nonlinear fashion. We calculated the nonlinear corrections
for several reactor designs spanning thermal neutron fluxes from
$\mathcal{O}(10^{12}\,\mathrm{n/cm^2/sec})$ to
$\mathcal{O}(10^{15}\,\mathrm{n/cm^2/sec})$, discovering a nonlinear
neutrino excess as large as $\sim 1\%$. The resulting nonlinear
nuclide production is negligible for short irradiation times less than
30\,d, but much larger for multi-batch commercial reactors, which can
reach large burn-up values.  This result indicates that special
attention should be given to the nonlinear correction in future
neutrino experiments, requiring detailed reactor simulations to
correctly predict this excess.

\acknowledgements We would like to thank A.~Hayes for the many
discussions on nonlinear nuclide production in a reactor as well as
D.~Ilas and the ORNL team behind the maintenance and verification of
SCALE. This work was supported by the U.S. Department of
Energy under award \protect{DE-SC0013632}.

\bibliographystyle{apsrev} \bibliography{NonLinearity.bib}

\end{document}